\input epsf.tex
\documentclass{article}
\usepackage{fullpage,doublespace,epsfig}
\author{Srinath Cheluvaraja}
\title{Thick vortices in $SU(2)$ lattice gauge theory}
\begin{document}
\def \beq{\begin{equation}}
\def \eeq{\end{equation}}
\def \ps{\psi}
\def \pb{\bar \psi}
\def \gi{\gamma_{i}}
\def \go{\gamma_{0}}
\def \g5{\gamma_{5}}
\baselineskip=24pt
\vspace{1.5cm}
\begin{center}
\bf{ Thick vortices in SU(2) lattice gauge theory}
\end{center}
\vspace{1.0cm}
\noindent
\begin{center}
\rm{Srinath Cheluvaraja} \\
\it{Center for Computational Biology and Bioinformatics, University of Pittsburgh,
Pittsburgh, PA, 15261}\\
\end{center}

\noindent{\bf{ABSTRACT}}\\
Three dimensional $SU(2)$ lattice gauge theory is studied after eliminating thin monopoles
and the smallest thick
monopoles. Kinematically this constraint 
allows the formation of thick vortex loops which produce $Z(2)$
fluctuations at longer length scales.
The thick
vortex loops are identified in a three dimensional simulation. A
condensate of thick vortices persists even after the
thin vortices have all disappeared. The thick vortices decouple
at a slightly lower temperature (higher $\beta$) than 
the thin vortices and drive a phase transition.
\rm
\vspace{0.5cm}
\begin{flushleft}
\end{flushleft}

\newpage 

The role of thick vortices as a mechanism for
confinement in $SU(2)$ LGT has been extensively discussed in the last
few years \cite{latts}.
Thick
vortices are  configurations analogous to domain walls in ferromagnets with a
continuous symmetry--they are like thick Peierls contours.
These vortices can reduce their free energy by increasing the size of
their core \cite{mack}. However, the competition between the energy and the entropy
prevents them from spreading indefinitely and a physical size emerges.
At distances beyond this physical size an effective theory of vortices
can be used to explain the long distance properties of the gauge theory.
In pure $SU(2)$ gauge theories vortices are defined with respect to the
center of the gauge group and
have a $Z(2)$ magnetic
charge. On the lattice,
thick vortices should be distinguished from thin vortices.
Thin vortices are like the vortices in a $Z(2)$ gauge theory--
their core has a thickness of only one lattice spacing and they are
suppressed by the plaquette term at large $\beta $.
A thin vortex consists of a set
of plaquettes with negative trace. 
It is
well known that a dilute gas of such thin vortices produces an area law for
the Wilson loop.
Because of flux conservation such configurations can form 
(in three dimensions) closed
loops or they can end at isolated points which absorb the vortex flux. 
These isolated
points are called $Z(2)$ monopoles.
The thin $Z(2)$ monopole density 
on a cube $c$ is
given by
\beq
\rho_{1}(c)=(1/2)(1-sign(\prod_{p \in \partial c} trU(p)))
\quad .
\eeq
In the above definition the product is taken over all the plaquettes
bordering a cube.
By definition, the monopole density is defined modulo $2$. Now we turn to
thick vortices.
Thick vortices can be identified as sets of $dXd$ loop variables
($d \ge 2$) having negative trace.
Unlike thin vortices the plaquettes in a thick vortex configuration
always have a positive sign. It is clear that such configurations are
only possible if the gauge group is continuous. A gas of such thick
vortices will also disorder Wilson loops like thin vortices.
Analogously, a thick vortex can end in a thick $Z(2)$ monopole.
Its density (in three space-time dimensions) is given by
\beq
\rho_{d}(c_{d})=(1/2)(1-sign(\prod_{d \in \partial c_{d}} trU(d)))
\quad ,
\eeq
where the product is now taken over all $dXd$ loops bordering a cube 
$c_{d}$ of
side $d$. This density is again defined modulo $2$.
The subscript $d$ indicates that the density can be defined on any
3 dimensional cube of side $d$.
Thus, in three space-time dimensions thick monopoles
have a finite size.
They are
monopoles because they
violate the Bianchi identity.
Densities of thin and thick vortices can be defined by
counting the number of plaquettes and dXd loops respectively
 with negative trace.
The interplay between monopoles
and vortices prevents an understanding of the gauge theory exclusively
in terms of one or the other. It should be clear that
a whole
hierarchy of such monopoles and vortices can be defined with different thicknesses.
Unlike the thin monopoles which are suppressed by the plaquette
action, the thick
monopoles cost lesser energy because their energy is spread over a finite
region.
It was proposed  in \cite{mack} that at low temperatures (larger $\beta$)
vortices and monopoles of larger cross-section will dominate
in determining long distance properties. These ideas were encapsulated
in the effective $Z(2)$ theory of confinement 
\cite{effz2} where it was argued that
the long distance properties of
$SU(2)$ theory becomes analogous to those of a
$Z(2)$ theory with a running coupling $\beta(d)$ ( $\beta(d) \rightarrow
\infty \  as \  d \rightarrow \infty$). As $\beta$ increases the effective
lattice spacing decreases and a thicker vortex will have the same size
in physical units. If these vortices are condensed
 in the large $\beta$ limit,
the continuum limit
at distances greater than the vortex size can be described by expanding
around a vortex condensate.
The above ideas summarise the
effective $Z(2)$ theory of confinement and it is an open question whether
this scenario is infact realised.

We should mention that there have been different approaches 
and results in the study of vortices
in non-abelian gauge theories. The free energy of a thick vortex was
calculated in \cite{kt} using twisted boundary conditions
and it was shown to be finite. 
Another approach to defining vortices is by gauge fixing
\cite{proj1} which leaves a $Z(2)$ symmetry intact.
Vortices are defined in the gauge fixed theory just as in a $Z(2)$
LGT.
Studies of vortex
excitations in this gauge have led to the observation of center dominance--
the ungauged degrees (the center degrees of freedom) are sufficient
to reproduce the string tension and some other non-perturbative quantities.
More discussion of this approach can be found in \cite{proj2,proj3,proj4}.
Another approach developed in \cite{kovtom} 
and pursued further in \cite{hay}  is to study monopoles
and vortices using the
$Z(2)XSO(3)$ decomposition of the $SU(2)$ LGT. In this
approach the role of the Wilson loop as a vortex counter is quite
transparent and it is capable of handling thin and thick
monopoles and vortices. 

The presence of thick monopoles prevents the thick vortices from forming
closed loops and makes it difficult to study their effects separately on
the lattice. 
If the thick monopoles are removed it might be possible
for us to see if thick vortices are
present at all and then study their effects.
With this aim in mind we
study the $SU(2)$ theory after eliminating the thin and the smallest
thick monopoles. This system can be regarded as a generalisation of
the Mack-Petkova (MP) model \cite{mack} where only thin monopoles are suppressed.
In the MP model the elimination of thin monopoles increases the entropy
of the vortex degrees of freedom at strong coupling and leads to a Z(2)
phase transition driven by thin vortices. We would like to see if the
elimination of the thick monopoles leads to a separate transition driven
by thick vortices.
Just as in the Mack-Petkova model the elimination of the thin and thick
monopoles changes only the short distance properties of the gauge theory
and long distance properties like confinement will remain unaffected. This
means that the physical continuum limit will still be at $\beta \rightarrow
\infty$ but we hope to get a hint of the role of thick vortices
at a larger value of $\beta$.

The model was simulated by introducing two chemical potential
terms for the thin and thick monopole densities in addition to the
Wilson action.
The action used was
\beq
S=\frac{\beta}{2}\sum_{p}tr\ U(p) -\lambda_{1}\sum_{c_{1}}\rho_{1}(c_{1})
-\lambda_{2}\sum_{c_{2}}\rho_{2}(c_{2})
\quad .
\eeq
This model was studied in 3 dimensions mainly for reasons of
computational speed and also because vortex lines are easier to
identify than vortex sheets. 
Since in three dimensions the coupling
constant is dimensionful the lattice spacing decreases by two if we
increase $\beta$ by a factor of two.
$\lambda_{1},\lambda_{2}$ were chosen large so that the monopole
densities are very small ( in practice $\lambda_{1}=\lambda_{2}\approx 10$
resulted in identically zero densities for almost all thermalised 
configurations on
$8^3$ lattices).

The simulation of this model presented its own share of difficulties.
The environment of a single link is complicated because each
link touches four thin monopoles and eight thick monopoles. Different
strategies were attempted to simulate this model. Metropolis updating and
a combination of heat bath and metropolis were both tried but
metropolis updating was found to be more efficient provided the
table of $SU(2)$ elements is tuned  regularly to get a reasonable acceptance.
Unlike the pure $SU(2)$ theory the lattice does not thermalise very
quickly.
It was found that in the phase
where thick vortices are in abundance an ordered start (which has zero
density of monopoles and vortices)
took a
longer time to reach the equilibrium distribution. Likewise, in the phase
where thick vortices are absent a random start ( which has a large number
of thick vortices) took a longer time to reach the equilibrium
distribution. These metastabilities do not arise in the pure $SU(2)$
theory or the Mack-Petkova model and they appear to be linked to the
thick vortices present in this model.
Several approaches were tried to deal with this metastability like
starting from configurations "closer" to the equilibrium configuration
but the problem always remained. Nevertheless, with increased computational
effort consistent results can be and were obtained.
The simulation was performed in
three dimensions mainly for reasons of computational speed.
On the lattices used ($8^3$) it was found that simulating this model
is more time consuming than simulating the four dimensional pure 
$SU(2)$ model and hence we have confined most of our studies to
$8^3$ lattices although some preliminary results for $12^3$ lattices were
also performed.

\begin{figure}
\epsfig{file=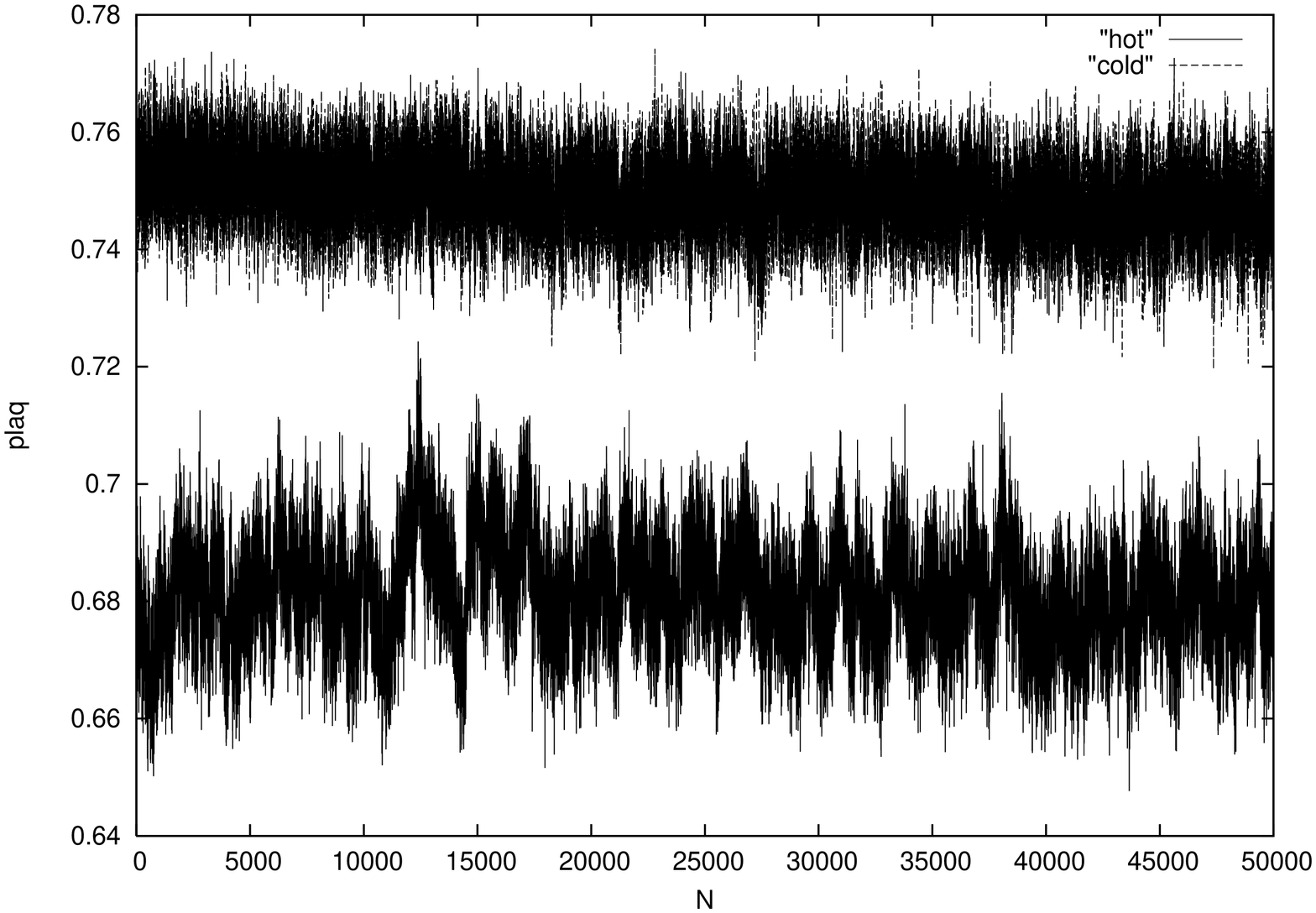,width=5cm,angle=-90}
\caption{ Metastability at $\beta=2.9$.  }
\label{meta}
\end{figure}

\begin{figure}
\epsfig{file=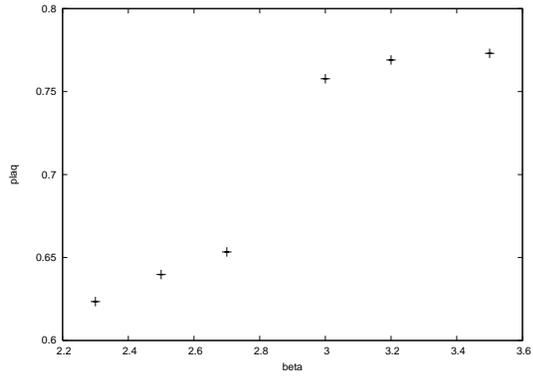,width=5cm,angle=-90}
\caption{Plaquette as a function of $\beta$.}
\label{plaq}
\end{figure}

\begin{figure}
\epsfig{file=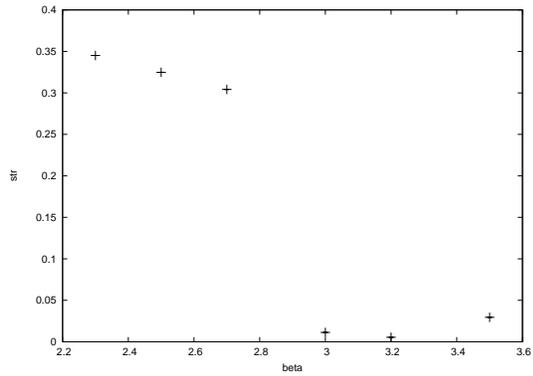,width=5cm,angle=-90}
\caption{Thick vortex density as a function of $\beta$. }
\label{vort}
\end{figure}

Several observables like the plaquette, the density of thin and thick
vortices, and Creutz ratios were studied in the simulations.
Around $\beta \approx 2.9$ very long metastabilities were observed
for the plaquette and $2X2$ Wilson loops. At this point ordered and
random configurations failed to converge and settled down to two
different values as shown in Fig.~\ref{meta}. As we go away from this point in either direction 
these metastabilities 
disappeared and the two starts resulted in identical Monte-Carlo
trajectories. For values of $\beta < 2.9$ a hot start was more efficient
in reaching the equilibrium configuration while the opposite was true for
$\beta > 2.9$. With these observations in mind we plotted the plaquette and
the density of thick vortices for different values of $\beta$. The jump
in the plaquette and the density of thick vortices also occurs at $\beta
\approx 2.9$ 
which is also the point where we observe two state behaviour.
The density of thick ($2X2$) vortices drops abruptly to a very small
value at the same value of $\beta$ as can be seen by comparing 
Fig.~\ref{plaq} and Fig.~\ref{vort}. The results in the graphs were
obtained by doing simulations from $100,000$ to $500,000$ MC steps.

\begin{figure}
\epsfig{file=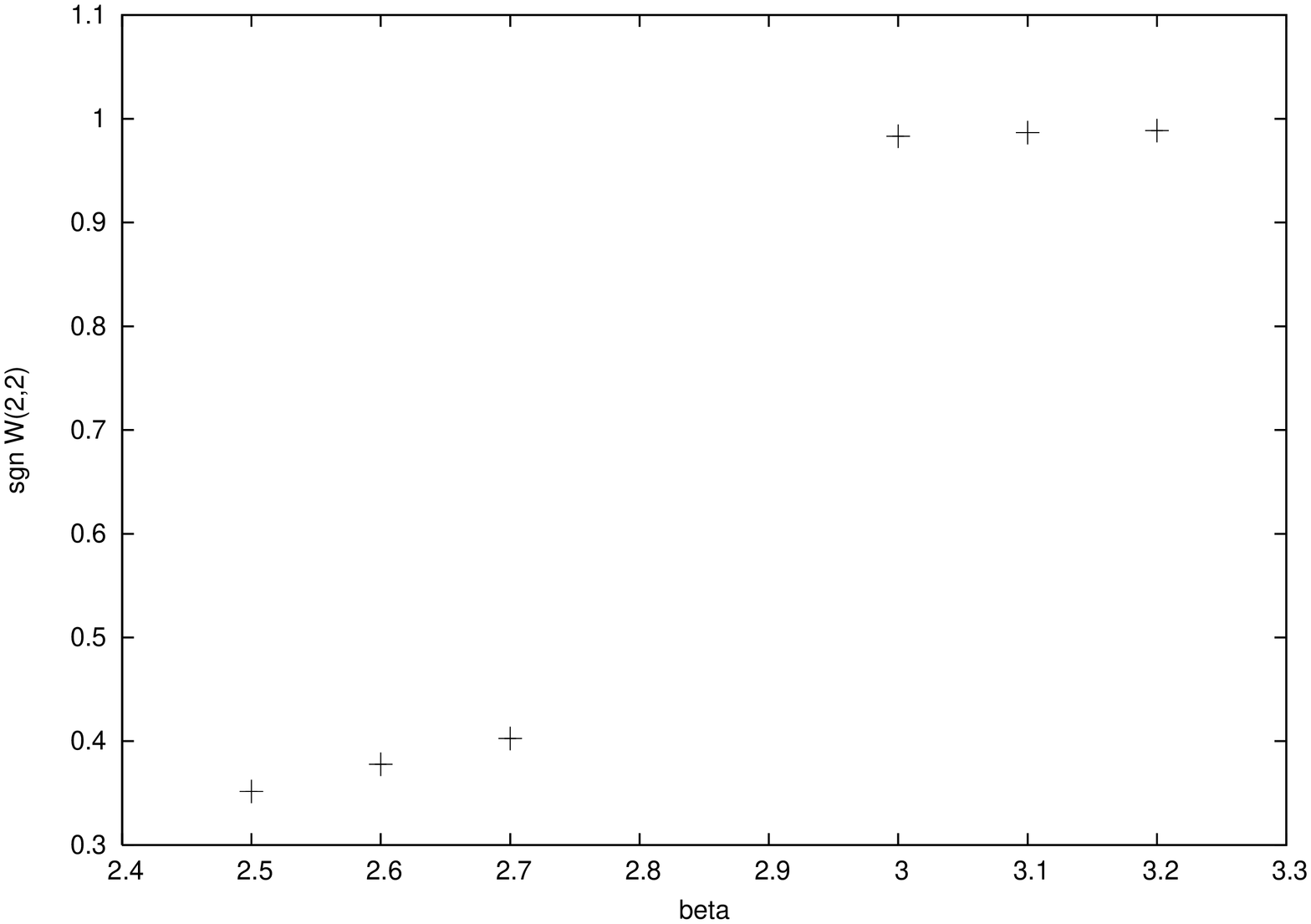,width=5cm,angle=-90}
\caption{Sign of the $2X2$ Wilson loop near the phase transition.}
\label{sign}
\end{figure}

It is well known that 3 dimensional $SU(2)$ LGT does not exhibit the
crossover found in the four-dimensional theory but instead has a very
smooth behaviour for the plaquette and other quantities. Elimination of only
thin monopoles leads to a slight jump in the plaquette 
(at  $\beta \approx 1.5$). The density of thin vortices drops to zero
at this point.
The interesting result of our simulations is that  
eliminating thick monopoles
causes another jump at a lower temperature (larger $\beta$).
In the vicinity of this jump very long relaxation times are observed
before a starting configuration reaches equilibrium.
The sign of the $2X2$ Wilson loop is a rapidly fluctuating quantity
near this transition and displays a more dramatic change than the
plaquette (compare Fig.~\ref{plaq} with Fig.~\ref{sign}).
It should be stressed that the thin vortices have
all disappeared near this transition and the sign of the plaquette is
a positive quantity in almost all configurations. $Z(2)$ monopoles of size
$d=1$ and $d=2$ are also absent. The dominant configurations are
those for which a $2X2$ Wilson loop has a negative sign such that all the
plaquettes inside the loop have a positive sign--in other words, configurations
which arise because the gauge group is continuous and non-abelian.
In the graphs we do not include the
points $\beta =2.8$ and $\beta=2.9$ because we observe strong metastability
near the phase transition. In three dimensions a doubling of $\beta$
corresponds to halving the lattice spacing and a thick vortex of twice
the lattice spacing would have the same physical size as a thin vortex
at the smaller $\beta$.

If $Z(2)$ fluctuations at longer length scales 
are present it is natural
to ask if these fluctuations are caused by thick vortices. 
Indeed they are!  Since vortex densities only provide information about
the number of vortex strings present but do not give any information of the
extent and size of the vortex loops we should track down the vortex loops
and get a measure of the different loop sizes present.
The set of $2X2$ loops with
negative sign must form closed loops by flux conservation with the
loops being defined on a lattice with twice the original lattice spacing.
A measurement
of the vortices on either side of the transition shows that they
form very long loops in one phase and smaller loops in the other
phase. 
The number of vortex loops $N(L)$
as a function of loop size $L$ at
$\beta =2.7$ on a $8^3$ lattice is given as follows:
\begin{center}
\begin{tabular}{|l|c|c|c|c|c|c|c|c|c|r|}
\hline
L & 8 & 12 & 16 & 20 & 28 & 32 & 36 & 44 & 60 & 64 \\
\hline
N(L) & 25 & 12 & 12 & 6 & 1 & 3 & 1 & 4 & 1 & 1 \\
\hline
\end{tabular}
\end{center}

The same on a $12^3$ lattice:
\begin{center}
\begin{tabular}{|l|c|c|c|c|c|c|c|c|c|c|c|c|c|c|c|c|c|c|c|c|c|r}
\hline
L & 8 & 12 & 16 & 20 & 24 & 28 & 32 & 36 & 40 & 48 & 52 & 56 & 60 \\
\hline
N(L) &52 & 14 & 3 & 9 & 7 & 2 & 2 & 1 & 2 & 2 & 2 & 4 & 3 \\
\hline
L & 72 & 80 & 88 & 96 & 100 & 104 & 108 & 116 & 128  \\
\cline{1-10}
N(L) & 1 & 2 & 1 & 1 & 2 & 1 & 1 & 1 & 1  \\
\cline{1-10}
\end{tabular}
\end{center}
The size of the largest loop increases with the lattice size and we expect
it to be of thermodynamic significance.
A comparison of the first few Creutz ratios in the phase where vortices dominate
with a nearby point
in the vortex less also shows a reduction by a factor of more than
four.

Determining the order of the phase transition needs
a systematic finite size scaling study of the features
observed on $8^3$ lattices. Studies on $12^3$ lattices show the same features
observed on $8^3$ lattices. In particular, the metastability observed
at $\beta=2.9$ persists on the larger lattice and even the size of the
discontinuity of the plaquette variable remains approximately the same.
A clear signature of a first order transition is a double peaked
histogram at the transition point.
Despite many improvements in our Monte-Carlo algorithm we have been unable
to observe any tunneling events and this has prevented us from making a
more detailed study of the transition. 
Since thin $Z(2)$ vortices drive a
second order transition in the 3 dimensional $Z(2)$ gauge theory, a first
order transition due to thick vortices would imply that fluctuations of
the non-centre degrees of freedom change the order of the transition.

The scaling properties of the $SU(2)$ theory only emerge in the
$\beta \rightarrow \infty $ limit and the thick vortices we have 
discussed are irrelevant in that limit as they decouple at
$\beta > 2.9$. However, at larger $\beta$  monopoles and vortices of
greater thickness $d(\beta)$ 
will still be present and can be in a condensed phase. This can also be
seen by studying the density of the vortices of thickness $3$ and greater.
These degrees of freedom are still important for $\beta > 2.9$ and they 
will decouple at a higher $\beta$ where
thicker vortices will become the dominant excitations.
As we approach the continuum limit
there will be a hierarchy of such effective $Z(2)$ theories operating at
larger and larger $\beta$ \cite{effz2}; the model studied here is just
the first member of the hierarchy.
Eliminating monopoles of greater thickness 
should in principle unravel
the entire hierarchy of $Z(2)$ like theories.
If vortices are relevant for
confinement the thick vortices will scale with $\beta$ so that a physical
size of the order of a fermi results.
This is one way in which the vortices of the $SU(2)$ LGT
can escape the phase transition seen in the $Z(2)$ LGT and always be in the
condensed phase.
It is not easy to
write down the effective $Z(2)$ theories for the different members of this
hierarchy but the effective $Z(2)$ theory will be in its strong coupling
phase \cite{kt} so that vortices are always abundant.
In \cite{brow} an effective $Z(2)$ theory
is derived for the $d=1$ case in
the strong coupling limit. Some properties of the effective
$Z(2)$ theory (in four dimensions) are discussed in \cite{effz2}. 

The main point of this investigation was to show that thick vortices
emerge as important excitations once the thick monopoles are suppressed.
The thick monopoles prevent the formation of long vortex loops and removing
them enhances the vortex degrees of freedom.
This leads to a jump in the value of the $2X2$ Wilson loop at large $\beta$.
The phase transition is driven by thick vortices, which
unlike the thin vortices, are fluctuations at longer distances. The
vortices responsible for these fluctuations can be explicitly identified. The
string tension decreases dramatically across this transition as signalled by
the change in the Creutz ratios. We have chosen to work in three dimensions
mainly for reasons of computational speed and the relative ease with which 
vortex lines can be identified. In four dimensions, monopoles and vortices will
be loops and sheets respectively.

\noindent

\newpage

\newpage
\begin{thebibliography}{99}
\bibitem{latts}{ Nucl. Phys. B (Proc. Suppl.) 106 (2002);Nucl. Phys. B (Proc. 
Suppl.) 119 (2003);Nucl. Phys. B (Proc. Suppl.) 129/130 (2004).}
\bibitem{mack}{G.~Mack and V.~B.~Petkova, Ann. of Physics {\bf 123} (1979) 442,
{\bf 125} (1980) 117;
E.~Tomboulis, Phys. Rev. {\bf D23} (1981) 2371.}
\bibitem{effz2}{G.~Mack, Phys. Rev. Lett. {\bf 45} (1980) 1578.}
\bibitem{kt}{T.~G.~Kovacs and E.~T.~Tomboulis, Phys. Rev. Lett. {\bf 85}
(2000) 704.}
\bibitem{kovtom}{T.~G.~Kovacs and E.~Tomboulis, Phys. Rev. {\bf D57} (1988) 4054.}
\bibitem{proj1}{L.~Del Debbio, M.~Faber, J.~Greensite, and S.~Olejenik, Phys.
Rev. {\bf D 55} (1997) 2298.}
\bibitem{proj2}{ M.~Faber, J.~Greensite, and S.~Olejenik, JHEP 9901:008,1999;
JHEP 9912:012,1999;hep-lat/9911006; hep-lat/9912002.}
\bibitem{proj3}{K.~Langfeld and H.~Reinhardt, Phys. Rev. {\bf D 55}, 7993 
(1997); K.~Langfeld, H.~Reinhardt, and O.~Tennert, Phys. Lett. {\bf 419} (1998)
317; M.~Engelhardt, K.~Langfeld, H.~Reinhardt, and O.~Tennert, Phys. Lett. {\bf 431} (1998) 141; {\bf 452} (1999) 301; Phys. Rev. {\bf D 61} (2000) 054504.}
\bibitem{proj4}{P.~de Forcrand and M.~D'Elia, hep-lat/9907028;hep-lat/9909005.}
\bibitem{hay}{A.~Alexandru and R.~W.~Haymaker, hep-lat/0002031.}
\bibitem{others}{A.~Hart, B.~Lucini, Z.~Schram, and M.~Teper, hep-lat/0005010;
C.~Korthals Altes, A.~Michels, M.~Stephanov, and M.~Teper, Phys. Rev.{\bf D55},
(1997) 1047.}
\bibitem{brow}{R.~C.~Brower, D.~A.~Kessler, and H.~Levine, Nucl. Phys. {\bf
B205}[FS5] (1982) 77.}
\end{thebibliography}
\end{document}